\documentclass{article}

\usepackage{amssymb,amsfonts,amsmath,stmaryrd}
\usepackage{cite,enumerate,float,indentfirst}
\usepackage{color}
\usepackage{tikz}
\usetikzlibrary{arrows,snakes,backgrounds}

\def\be{\begin{eqnarray}}
\def\ee{\end{eqnarray}}
\def\nn{\nonumber}
\def\la{\langle}
\def\ra{\rangle}

\def\tr{{\rm tr}}
\def\Tr{{\rm Tr}}

\definecolor{red}{rgb}{1,0,0}
\definecolor{orange}{rgb}{1,0.5,0}
\definecolor{violet}{rgb}{0.7,0,1}

\hoffset= - 1.0in         

\begin{document}

\hfill MIPT/TH-08/23

\hfill ITEP/TH-09/23

\hfill IITP/TH-07/23

\bigskip

\bigskip

\bigskip

\centerline{\Large{On information paradox and the fate of black holes
}}

\bigskip

\centerline{\bf A.Morozov}

\bigskip

\centerline{\it MIPT, ITEP \& IITP, Moscow, Russia}

\bigskip

\centerline{ABSTRACT}

\bigskip

{\footnotesize
A sketchy review of the "island" paradigm in black hole evaporation theory,
which actually brings us back to the old idea that interior of black hole
decouples from our universe after Page time, so that Hawking radiation
is entangled with emerging new universe,
thus leaving no room for the information paradox.
Instead this provides a self-consistent description of multiverse,
where every black hole in a parent universe is a white hole -- the origin --
of a new one.

}

\bigskip

\bigskip

Information paradox is a long-standing puzzle in the theory of black hole (BH) evaporation \cite{evapo}.
Recently it attracted a new attention \cite{islands} because of the new source of evidence,
provided by AdS/CFT-correspondence.
Surprisingly or not, it brought some of us back to the old point of view,
that there is no paradox at all.
This is a brief review of what we understand about the present state of affairs in this field.

The essence of the paradox  \cite{Hawking}
is that starting from a {\it pure} quantum state we can go through
a formation of black hole to its complete evaporation, leaving one with a thermal ensemble
for emitted Hawking radiation -- what drastically contradicts unitarity of quantum evolution,
which can not convert {\it pure} states into {\it mixed}.

There was a variety of radical attempts to resolve the paradox,
going deep into different projects of quantum gravity.
However, in full accordance with Okkam's razor principle, one does not need to go so far --
the paradox is resolved at the same level as it is formulated,
i.e. without any reference to quantum gravity.

\section{The basic scenario}

The old/new resolution is that the final state is very different from above assumption.
Namely, at the Page time \cite{Page} a part of the black hole interior
starts to decouple from our universe, forming the Wheeler's "bag of gold" \cite{Wheeler}
or a wormhole to a new universe.
In modern language this is called an {\it island} inside the black hole,
the domain which can not be seen in our universe,
even by entropic considerations.
At the very end, the "observable" part of black hole fully disappears (evaporates),
and thus entanglement entropy of this "nothing" with radiation vanishes --
and the true entanglement is between {\it thermal} Hawking radiation {\bf in} our universe
and the island {\bf beyond} it.
Together they continue to be in the {\it pure} state,
but separately they look {\it mixed}.
The archetypical example is the {\it pure} thermodouble state
\be
\rho_{_{TD}} = \left|\Psi_{_{TD}}\right>\left<\Psi_{_{TD}}\right|
\ee
with
\be
\left|\Psi_{_{TD}}\right> = \sum_i e^{-E_i/2T}\, \left|i\right>_{_I}\otimes \left|i\right>_{_R}
\ee
and {\it mixed} states
\be
\rho_{_R}   = \Tr_{_I} \,\rho_{_{TD}} = \sum_i e^{-E_i/T} \, \left|i\right>_{_R}\left<i\right|_{_R}
\ee
\be
\rho_{_I}  = \Tr_{_R} \,\rho_{_{TD}} = \sum_i e^{-E_i/T} \, \left|i\right>_{_I}\left<i\right|_{_I}
\ee


\begin{figure}[h]

\begin{picture}(250,150)(-230,-50)

\qbezier(-50,50)(-50,80)(0,80)  \qbezier(50,50)(50,80)(0,80)
\qbezier(-20,20)(-50,20)(-50,50) \qbezier(20,20)(50,20)(50,50)
\qbezier(-5,10)(-5,20)(-20,20) \qbezier(5,10)(5,20)(20,20)
\qbezier(-15,-5)(-5,0)(-5,10) \qbezier(15,-5)(5,0)(5,10)
\qbezier(-15,-5)(-35,-10)(-100,-30) \qbezier(15,-5)(35,-10)(100,-30)

\put(-20,-35){\mbox{\text{Universe}}}
\put(-7,-10){\mbox{\text{BH}}}
\put(15,5){\mbox{\text{horizon}}}
\put(-47,45){\mbox{\text{decoupled bag of gold}}}

\end{picture}

\caption
{{\footnotesize
Wheeler's "bag of gold" is a solution to Einstein equations
where interior of the black hole is substituted by geometry
with a fixed area (equal to horizon area)
and arbitrarily large volume.
We need a quasiclassical treatment of this kind of geometries,
when they coexist with Hawking radiation around in the main
part of the Universe, outside the Schwarzschild radius.
}}

\end{figure}
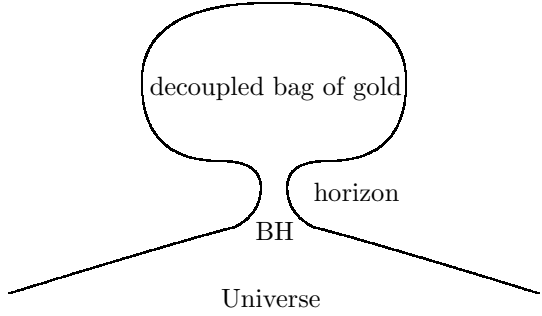

This interpretation of BH evaporation implies that every black hole in a given universe
is a white-hole source of a new one,
thus providing a tree model of multiverse, see s.\ref{multi} below.
Each decoupling {\it bag of gold = black hole for us and a white-hole for some emerging new world}
has all the quantum numbers
(charges), associated with gauge symmetries, vanishing,
but can have non-trivial entanglement with the parent universe
and also non-trivial values of ungauged charges, like baryon number.
what can provide  new options to resolve the  baryon asymmetry problem.

\newpage

\section{Relevant theses}

The following comments are very simple, still they can be useful for thinking about the information paradox.

\begin{itemize}

\item{
Entropy growth is consistent with unitary evolution.

}

{
	Of course, if $\rho$ represents the state of the whole system, then
	\be
		\tr U\rho U^\dagger \log (U\rho U^\dagger) = \tr \rho\log\rho
	\ee
i.e. the entropy does not change.
However, the entropy of some part 1 of the system $1 \cup 2$, which is described by $\rho_1 = \tr_{\!2} \rho$,
is not conserved:
	\begin{align}
		\rho' &= U\rho U^\dagger, & \rho_1' &= \tr_{\!2}\, \rho' \nn\\
		\tr \rho' \log \rho' &= \tr \rho \log \rho, & \tr_{\!1} \rho_1' \log \rho_1' &\ne \tr_{\!1} \rho_1 \log \rho_1.
	\end{align}
	Thus, entropy growth of a part is consistent with unitary evolution of the whole system.
}

\begin{itemize}
\item{
In particular, burning a fire is consistent with unitary evolution.\\
{
	The whole system containing the fire and its radiation is assumed to evolve unitary.
However, when we trace out information that runs away with the radiation,
the entropy of remaining system (fire) can -- and it does so.
}
}
\end{itemize}

\item{Entropies of both subsystems $1$ and $2$ can increase simultaneously,
while entire system remains in the pure state, i.e. has vanishing entropy:
see (\ref{S1=S2}) below.
}

\item{Entropy and entanglement entropy are intimately related notions.
\begin{itemize}
\item{Entropy of a subsystem can be defined only as entanglement entropy. \\
{
	If an incompleteness of our knowledge about a subsystem is only due to absence of knowledge about complementary subsystem, the entropy of subsystem coincides with entanglement entropy.
	In general, entanglement entropy is a lower bound on entropy of subsystem.\\
}
}
\item{Ordinary entropy is equal to entanglement entropy {\it with} a pure state.\\
{
	Alternatively it may seem that ``entanglement entropy {\it with} a pure state'' is an ill defined object,
    and entanglement entropy is well-defined only in the case when the state of whole system is pure.
	For example, tensor product of the density matrix of a mixed state and that of a pure state is not pure:
	\be
		\tilde \rho = \rho \otimes \big(|\psi\ra\la\psi|\big), \qquad \tr \tilde\rho \log\tilde\rho = \tr \rho \log\rho \ne 0
	\ee
	As a consequence entropy of reduced density matrices are not equal to each other
	\be
		\rho_1 = \tr_{\!2} \tilde \rho = \rho, \qquad \tr_{\!1}\rho_1 \log \rho_1 \ne 0 \nn\\
		\rho_2 = \tr_{\!1} \tilde \rho = |\psi\ra\la\psi|, \qquad \tr_{\!2}\rho_2 \log \rho_2 = 0
	\ee
}}
\item{If entire system is in pure state, entanglement entropies of its parts
sum to zero. \\
{
	This depends on the meaning of ``sum to zero''. The situation is the following
	\begin{enumerate}
		\item Take the whole system in pure state, whose parts 1 and 2 are (possibly) entangled
		\be
			|\psi\ra = \sum_i c_i |i\ra_1 |\tilde i\ra_2
		\ee
		\item Now, forget about (for example) subsystem 2, so that we see the subsystem 1 in the state
		\be
			\rho_1 = \tr_{\!2}|\psi\ra \la\psi| = \sum |c_i|^2 |i\ra_1 \la i|_1
		\ee
		Incompleteness of our knowledge about subsystem 1 results into the entropy
		\be
			S(\rho_1) = - \tr_{\!1} \rho_1 \log \rho_1
		\ee
		\item Moreover, incompleteness of our knowledge about the subsystem 2 after forgetting about the subsystem 1 equals to incompleteness of our knowledge about the subsystem 1 after forgetting about the subsystem 2
		\be
			S(\rho_1) = S(\rho_2)
        \label{S1=S2}
		\ee
		\item Thus, if one remembers everything about complement subsystem the entropies of the subsystem ``collapse into zero''.
	\end{enumerate}
	\textbf{Note:} The only role of the word ``entanglement'' before ``entropy'' is to emphasize that the \emph{nature}
of incompleteness is that the whole system state was entangled state, namely it's subsystems 1 and 2 were somehow correlated.
}
}
\end{itemize}
}

\item{
Complement of the thermal state to a pure one should be at least as big:
\be
\left(\sum_{i=1}^N e^{-\beta E_i}|i><i|\right) \otimes
\left(\sum_{a=1}^n u_a |a><a|\right) =
|\psi><\psi|  \ \Longrightarrow \ n\geq N
\ee
{
	The easiest way to purify a mixed state
	\be
		\rho =\sum_i p_i |i\ra \la i|, \qquad \la i | j \ra = \delta_{ij}  \label{dm_mixed}
	\ee
	is the following. Take the double-copied Hilbert space $\mathcal H \otimes \mathcal H$ and consider the pure state
	\be
		|\psi\ra = \sum_i \sqrt{p_i} \; |i\ra \otimes |i \ra
	\ee
	Tracing of corresponding density matrix $\rho_\mathrm{pure} = |\psi\ra \la \psi|$ over the second copy of the Hilbert space leads to the initial density matrix
	\be
		\tr_\mathrm{\!copy} \, \rho_\mathrm{pure} = \sum_{i,j} \sqrt{p_i p_j} \; \la j|i \ra \; |i\ra \la j| = \sum_i p_i |i\ra \la i| = \rho
	\ee
	In the case of thermodynamic Gibbs distribution density matrix
	\be
		\rho_\mathrm{th} = \sum_i e^{-\beta E_i} |i\ra \la i|
	\ee
	the purified state is so-called thermofield double state
	\be
		|TFD\ra = \sum_i e^{-\beta E_i/ 2}  |i\ra_1 \otimes |i \ra_2, \qquad \tr_{\!2}  |TFD\ra \la TFD| = \rho_\mathrm{th}
	\ee
	which is said to be holographic dual to eternal AdS-Schwarzschild black hole.
	
	In the case, where some of probabilities in (\ref{dm_mixed}) vanish, dimension of additional Hilbert space, used for the state purification, can be reduced. Indeed, introducing Hilbert space
	\be
		\tilde {\mathcal H} = \operatorname{Span}(|\tilde i\ra), \quad \tilde i = 1, \ldots (\text{\# of $p_i\ne0$})
	\ee
	one can purify (\ref{dm_mixed}) as
	\be
		|\psi\ra = \sum_{i: \,p_i\ne 0} \sqrt{p_i} \; |i\ra \otimes |\tilde i \ra.
	\ee
}
}

\item{
Evaporation of one-half of the black hole content defines the Page time --
and this is the last moment at which its interior (back of gold or "island")
{\it should} decouple.
Of course, this can happen earlier, but this depends on the still unknown
details of the black hole evolution.
It is even unclear what are the factors, which can affect the answer to this question.
}

\item{Horizon area calculates only a "part" of the states inside the black hole.
These are the states "visible" from the outside and their number decreases to zero
in the process of evaporation.
}

\item{
Essential part of evaporating black hole decouples from our universe at around Page time
(as a "bag of gold" or as a wormhole to somewhere else).
This can be interpreted as formation of an {\it island} \cite{islands}.
}

\item{
Their can be various assumptions about the fate of evaporating black hole,
we mention just a few:
\begin{itemize}
\item{
It partly disappears from our universe and can become an origin of a new one.

}
\item{
Absorbed {\it energy} is fully radiated back to us,
new universe is born with net zero energy.
}
\item{
Absorbed information is partly transferred to the new universe.
}

\item{

Perhaps, it is born with non-vanishing information
{\it or}
it is born, with vanishing info, but entangled with our universe.

}

\item{Entropy of our universe gets increased after evaporation.}

\item{A new universe is born with vanishing entropy.}

\end{itemize}
It is not clear which of these theses are fully correct and what are the factors
which can regulate the possible deviations.
}

\end{itemize}

\section{History of the multiverse
\label{multi}}

Since in this scenario Hawking radiated is only a part
(at most half) of what falls into the black hole,
its  interior remains non-trivial and decouples from our Universe.
Afterwards it lives its own life, i.e. becomes a separate
new universe.
If its history is similar to ours, we get a typical
multiverse pattern:

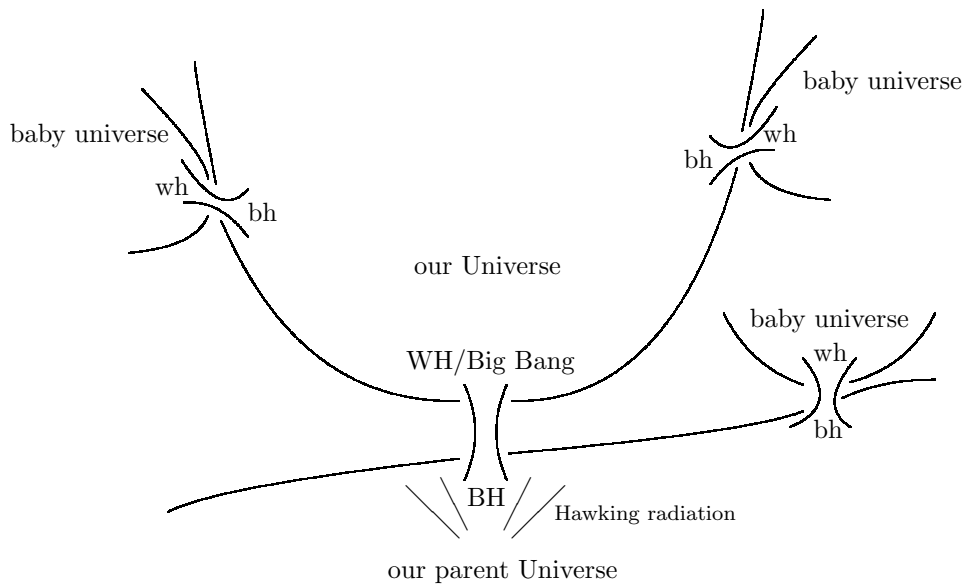
\begin{figure}[h]
\begin{picture}(200,250)(-200,-80)

\put(-5,20){
\qbezier(114,77)(100,78)(90,64) \qbezier(115,93)(100,70)(90,82)
\qbezier(102,84)(110,125)(110,130)
\qbezier(105,72)(110,60)(135,58) \qbezier(105,86)(106,95)(130,120)
\put(125,100){\mbox{\text{baby universe}}}
\put(110,80 ){\mbox{\text{wh}}}
\put(80,70){\mbox{\text{bh}}}
}

\qbezier(-114,77)(-100,78)(-90,64) \qbezier(-115,93)(-100,70)(-90,82)
\qbezier(-102,84)(-110,125)(-110,130)
\qbezier(-105,72)(-110,60)(-135,58) \qbezier(-105,86)(-106,95)(-130,120)
\put(-180,100){\mbox{\text{baby universe}}}
\put(-125,80 ){\mbox{\text{wh}}}
\put(-90,70){\mbox{\text{bh}}}

\qbezier(10,2)(70,0)  (95,90)\qbezier(-10,2)(-70,0)  (- 100,70)
\put(-27,50){\mbox{\text{our Universe}}}

\put(-30,14){\mbox{\text{WH/Big Bang}}}
\qbezier(-8,8)(0,-10)(-8,-28) \qbezier(8,8)(0,-10)(8,-28)
\put(-7,-38){\mbox{\text{BH}}}

\qbezier(9,-18)(100,-10)  (120,-2) \qbezier(-10,-20)(-100,-30)  (-120,-40)
\qbezier(170,10)(150,10)  (135,3)
\put(-37,-65){\mbox{\text{our parent Universe}}}

\put(100,30){\mbox{\text{baby universe}}}
\put(124,17){\mbox{\text{wh}}}
\qbezier(120,18)(135,0)(115,-8) \qbezier(140,18)(125,0)(138,-8)
\qbezier(120,8)(100,15)(90,35)  \qbezier(138,9)(160,15)(170,35)
\put(124,-12){\mbox{\text{bh}}}

\put(-10,-50){\line(-1,1){20}}
\put(-7,-47){\line(-1,2){10}}
\put(10,-50){\line(1,1){20}}
\put(7,-47){\line(1,2){10}}

\put(26,-43){\mbox{\footnotesize Hawking\ radiation}}

\end{picture}

\caption{{\footnotesize Schematic picture of the history of Universe.
It is created as a white hole from a black hole in the parent Universe.
Other black holes give rise to other baby universes --
some of them are sisters of our, others are descendants and so on.
Each black hole is surrounded by Hawking radiation, which stays in its own
universe, while the other content of the black hole decouples from it
at around the Page time and gives birth to a new baby universe.
Inside the new universe the origin look like a white hole or a Big Bang.
}}

\end{figure}

\section{Open questions}

It is an almost unavoidable prediction of General Relativity that every universe
has a simple white hole at its origin (big bang)
and many black holes, formed during its evolution —
with each black hole giving birth to a new baby/descendant universe
with its own fate.

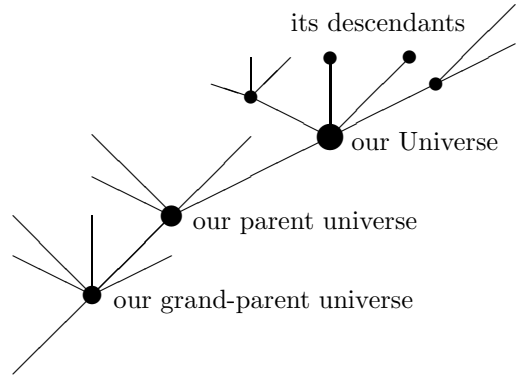
\begin{figure}[h]
\begin{picture}(200,130)(-200,-55)

\put(60,30){\circle*{10}}
\put(60,30){\line(-2,1){30}}
\put(60,30){\line(2,1){40}}
\put(60,30){\line(1,1){30}}
\put(60,30){\line(0,1){30}}

\put(90,60){\circle*{5}}
\put(30,45){\circle*{5}}
\put(60,60){\circle*{5}}
\put(100,50){\circle*{5}}

\put(30,45){\line(-3,1){15}}
\put(30,45){\line(0,1){15}}
\put(30,45){\line(1,1){15}}
\put(100,50){\line(1,1){30}}
\put(100,50){\line(2,1){30}}

\put(0,0){\circle*{8}}
\put(0,0){\line(-2,1){30}}
\put(0,0){\line(2,1){60}}
\put(0,0){\line(1,1){30}}
\put(0,1){\line(-1,1){30}}
\put(0,0){\line(-1,-1){30}}

\put(68,25){\mbox{\text{our Universe}}}

\put(8,-5){\mbox{\text{our parent universe}}}

\put(-22,-35){\mbox{\text{our grand-parent universe}}}

\put(45,70){\mbox{\text{its descendants}}}

\put(-30,-30){\circle*{7}}
\put(-30,-30){\line(-2,1){30}}
\put(-30,-30){\line(2,1){30}}
\put(-30,-30){\line(0,1){30}}
\put(-30,-30){\line(1,1){30}}
\put(-30,-30){\line(-1,1){30}}
\put(-30,-30){\line(-1,-1){30}}
 
\end{picture}

\caption{{\footnotesize Another schematic picture of the multiverse.
Black dots show ancestors of our Universes,
lines are the black-white hole transitions between the universes.
Many other universes, created from various black holes
and their further descendants are not shown.
Each universe has a single white-hole origin and many black holes,
which give rise to many descendent baby-universes.
Some of them can collapse before giving birth to anything new,
some can have a lot of descendants.
}}
\end{figure}

This {\it prediction} should be distinguished from an additional {\it possibility} that
there can be wormholes, some looking like black-white hole pairs,
which can connect different universes or different regions in the same universe —
these are still only speculations, depending on various assumptions
beyond General Relativity {\it per se}.

The main technical problem is the quasiclassical description of Wheeler's bag-of-gold
geometries, which would provide a consistent description of
Hawking radiation and new-born baby-universes.
In particular, this should answer the question, which part of matter,
falling into the black hole, is actually radiated
(at most half, but it can actually be much smaller)
and what are the parameters, which regulate this fraction.
We hope that more attention will be paid to critical analysis
of this kind of naive scenario for black hole evaporation
and this will finally lead to resolution of paradoxes and
to healthy theories of multiverse dynamics.

\section*{Acknowledgements}

I am indebted for valuable discussions and comments to Nikita Kolganov.

My work is supported by the Russian Science Foundation (Grant No.21-12-00400).

 \end{document}